\documentclass[10pt]{article}

\begin{document}

\title{Late time cosmic acceleration from vacuum Brans-Dicke theory in $5D$}
\author{J. Ponce de Leon\thanks{E-Mail:
jpdel@ltp.upr.clu.edu, jpdel1@hotmail.com}  \\Laboratory of Theoretical Physics, 
Department of Physics\\ 
University of Puerto Rico, P.O. Box 23343,  
San Juan,\\ PR 00931, USA}
\date{Revised version. February, 2010}

\maketitle
\begin{abstract}
We  show that the scalar-vacuum Brans-Dicke equations in $5D$ are equivalent to  Brans-Dicke theory  in $4D$ with a self-interacting potential and an effective  matter field. The cosmological implication, in the context of FRW models,  is that the observed accelerated expansion of the universe comes naturally from the condition that the scalar field is not a ghost, i.e.,  $\omega > - 3/2$.  We find an effective matter-dominated $4D$ universe which shows accelerated expansion if  $- 3/2 < \omega < - 1$. We study the question of whether accelerated expansion can be  made compatible with large values of $\omega$, within the framework  of a  $5D$ scalar-vacuum Brans-Dicke theory with  variable,  instead of constant, parameter $\omega$.  In this framework, and based on a general class of solutions of the field equations, we demonstrate  that   accelerated expansion is incompatible with large values of $\omega$.

\end{abstract}

PACS number(s): 04.20.Jb, 04.50.+h, 98.80.Es

\newpage

\section{Introduction}
Currently,  there is a general agreement among cosmologists that the expansion of the universe is speeding up, instead of slowing down. Evidence  in favor of this  is provided by observations of high-redshift supernovae Ia \cite{Riess}-\cite{Tonry}, the cosmic microwave background and galaxy power  spectra \cite{Lee}-\cite{Sievers}.

Since the gravity of both matter and radiation is attractive, an accelerated expansion requires 
either modified Einstein equations or, in the context of general relativity, the presence  of a mysterious form of matter (called {\it dark energy}), which accounts for  $70\%$ of the total content of the universe and  remains unclustered on all scales where gravitational clustering of ordinary matter  is seen.

Within the context of  general relativity the simplest candidate for dark energy is the cosmological constant $\Lambda$ \cite{Peebles}-\cite{Padmanabhan0}.  However, to avoid the  `cosmic coincidence problem'  \cite{Zlatev} many researchers  consider a time dependent cosmological term, or an evolving scalar field known as {\it quintessence} \cite{Zlatev}-\cite{Deustua}. 
The main drawback in these models is that the scalar fields are introduced  by `hand', without explaining the  origin  of the fields.

In recent years there has been a renewed interest in scalar-tensor theories of gravity as viable alternatives to general relativity. In particular, some researchers resort to Brans-Dicke theory (BD) to explain the accelerated expansion of the universe from fundamental physics.  The concept is  that the scalar field  in  BD could lead to cosmological acceleration. However, it turns out that this is so only in very particular cases: for values of the coupling parameter $\omega$ in the range $- 2 \leq \omega < - 3/2$, which violate the energy condition on the scalar field;  for  $\omega $ varying  with time \cite{Pavon}; when there is a
scalar potential, which in turn is  added on an {\it ad hoc} basis \cite{Sen}. Further models include the so-called `chameleon fields' that allow the scalar field to interact with matter (see \cite{Das} and references therein). 
A different approach comes from Brans-Dicke theory in $5D$. It has recently been shown \cite{Li} that the reduced theory in $4D$ is equivalent to ordinary general relativity with two scalar fields, which can be used to explain the present accelerated expansion of the universe. In our view the main deficiency, or controversial feature,  in this approach is the {\it ad hoc} introduction of matter in $5D$. 

The present work is motivated by \cite{Li}. Here,  our object  is to study the question of whether we can obtain the observed accelerated expansion in $4D$ from Brans-Dicke theory in $5D$ {\it without} assuming the existence of higher-dimensional matter.  
We show that the answer to this question is positive. We  find that the scalar-vacuum BD equations in $5D$ are equivalent to  BD in $4D$ with a self interacting potential and an effective  matter field. 
In our approach,  the presence of a non-vanishing potential is crucial to recover a general BD theory in $4D$; the attractive feature is that its shape is determined, up to a constant, by the reduction procedure. 
It turns out that the accelerated expansion of an effective  dust  universe comes naturally from the condition $\omega > - 3/2$, in the region $-  3/2 < \omega < - 1$. This range of $\omega$ is consistent  with the one obtained in \cite{Li}.

However, small values of $|\omega|$  are in sharp contradiction with  the solar system bound $\omega > 600$. Therefore, as a possible  way out of this problem, we consider a    $5D$ scalar-vacuum BD theory with {\it variable} $\omega$ and study the question of whether accelerated expansion is compatible with large values of $\omega$. In $4D$ this approach has been suggested by Banerjee and Pavon  \cite{Pavon}.   Our analysis indicates that the answer to our question is negative. Namely, late time accelerated expansion is incompatible with large values of $\omega$.

\section{Dimensional reduction of  Brans-Dicke theory in $5D$}

The Brans-Dicke theory of gravity in $5D$ is described by the action \cite{Li}
\begin{equation}
\label{action}
S_{(5)} = \int{d^5 x \sqrt{|\gamma^{(5)}|}\left[\phi R^{(5)} - \frac{\omega}{\phi}\gamma^{A B}\left(\nabla_{A}\phi\right)\left(\nabla_{B}\phi\right) 
\right]}  + 16 \pi \int{d^5 x \sqrt{|\gamma^{(5)}|}L_{m}^{(5)}}, 
\end{equation}
where $R^{(5)}$ is the curvature scalar associated with the $5D$ metric $\gamma_{A B}$; $\gamma^{(5)}$ is the determinant of $\gamma_{A B}$; $\phi$ is a scalar field; $\omega $ is a dimensionless coupling constant; $L_{m}^{(5)}$ represents the Lagrangian of the matter fields in $5D$ and does not depend on $\phi$.  We consider only scalar-vacuum configurations in $5D$, i.e., put $L_{m}^{(5)} = 0$.

The equations for the gravitational field in $5D$ derived from (\ref{action}) read 

\begin{equation}
\label{equation for the metric in 5D}
G_{AB}^{(5)} = R_{AB}^{(5)} - \frac{1}{2}\gamma_{AB}R^{(5)} = \frac{\omega}{\phi^2}\left[\left(\nabla_{A}\phi\right) \left(\nabla_{B}\phi\right) - \frac{1}{2}\gamma_{AB} \left(\nabla^{C}\phi\right)\left(\nabla_{C}\phi\right)\right] + \frac{1}{\phi}\left(\nabla_{A}\nabla_{B}\phi - \gamma_{AB} \nabla^2 \phi\right).
\end{equation}
where $\nabla^2 \equiv \nabla^{C}\nabla_{C}$.
 The field equation for the scalar field $\phi$ is determined by (\ref{action}) as
\begin{equation}
\label{dalambertian in 5D in terms of R5}
\frac{2 \omega}{\phi}\nabla^2 \phi - \frac{\omega}{\phi^2}\left(\nabla_{A}\phi\right)\left(\nabla^A \phi\right) + R^{(5)} = 0.
\end{equation}
Taking the trace of (\ref{equation for the metric in 5D}) we find 
\begin{equation}
\label{R5}
R^{(5)} =  \frac{\omega}{\phi^2}\left(\nabla_{A}\phi\right)\left(\nabla^{A}\phi\right) + \frac{8}{3\phi}\nabla^2\phi.
\end{equation}
Consequently, 
\begin{equation}
\label{equation for phi}
\nabla^2\phi = 0.
\end{equation}

\medskip

In this work we  use coordinates where the metric in $5D$ can be written as\footnote{Notation:  $x^{\mu} = (x^0, x^1, x^2, x^3)$ are the coordinates in $4D$ and $y$ is the coordinate along the extra dimension.  We use spacetime signature $(+, -, -, -)$, while $\epsilon = \pm 1$ allows for spacelike or timelike extra dimension. }  
\begin{equation}
\label{line element in 5D}
dS^2 = \gamma_{AB}d x^A d x^B = g_{\mu\nu}(x, y)d x^{\mu}d x^{\nu} + \epsilon \Phi^2(x, y) d y^2,
\end{equation}
in such a way that  our $4D$ spacetime can be recovered by going onto a hypersurface $\Sigma_{y}:y =  y_{0} = $ constant, which is orthogonal to the $5D$ unit vector
\begin{equation}
\label{unit vector n}
{\hat{n}}^{A} = \frac{\delta^{A}_{4}}{\Phi}, \;\;\;n_{A}n^{A} = \epsilon,
\end{equation}
along the extra dimension, and $g_{\mu\nu}$ can be interpreted as the metric of the spacetime. To maintain contact with \cite{Li} we assume that ${\hat{n}}^{A}$ is a Killing vector, which   in practice means that the metric coefficients in (\ref{line element in 5D}) only depend on $t$.

The effective field equations (FE) in $4D$ are obtained from dimensional reduction of  (\ref{equation for the metric in 5D}) and  (\ref{equation for phi}). 
To achieve such a reduction we note that 
\begin{eqnarray}
\nabla_{\mu}\nabla_{\nu}\phi &=& D_{\mu}D_{\nu}\phi, \nonumber \\
\nabla_{4}\nabla_{4}\phi &=&  \epsilon \Phi \left(D_{\alpha}\Phi\right)\left(D^{\alpha}\phi\right),\nonumber \\
 \nabla^2\phi &=& D^2 \phi + \frac{\left(D_{\alpha}\Phi\right)\left(D^{\alpha}\phi\right)}{\Phi}
\end{eqnarray}
where  $D_{\alpha}$ is the covariant derivative on $\Sigma_{y}$, which is calculated with $g_{\mu\nu}$, and $D^2 \equiv D^{\alpha}D_{\alpha}$. 

Using these expressions, the spacetime components ($A = \mu, B = \nu$) of the $5D$ field equations (\ref{equation for the metric in 5D}) can be written as 

\begin{eqnarray}
\label{spacetime components of G5}
G_{\mu\nu}^{(5)} =  - \frac{g_{\mu\nu}\left(D_{\alpha}\Phi\right)\left(D^{\alpha}\phi\right)}{\Phi \phi} + \frac{\omega}{\phi^2}\left[\left(D_{\mu}\phi\right) \left(D_{\nu}\phi\right) - \frac{1}{2}g_{\mu\nu} \left(D_{\alpha}\phi\right)\left(D^{\alpha}\phi\right)\right] + \frac{1}{\phi}\left(D_{\mu}D_{\nu}\phi - g_{\mu\nu} D^2\phi\right).
\end{eqnarray}

To construct the Einstein tensor in $4D$ we have to express $R_{\alpha\beta}^{(5)}$ and $R^{(5)}$ in terms of the corresponding $4D$ quantities.
The Ricci tensor $R^{(4)}_{\mu\nu}$ of the metric $g_{\mu\nu}$ and the scalar field $\Phi$ are related to the Ricci tensor $R_{AB}^{(5)}$ of $\gamma_{AB}$ by
\begin{eqnarray}
\label{dimensional reduction}
R_{\alpha\beta}^{(5)} &=& R_{\alpha\beta}^{(4)} - \frac{D_{\alpha}D_{\beta}\Phi}{\Phi}, \nonumber \\
R_{44}^{(5)} &=& - \epsilon \Phi D^2 \Phi.
\end{eqnarray}
From (\ref{equation for the metric in 5D})-(\ref{R5}) and the second equation in (\ref{dimensional reduction}) we obtain the $4$-dimensional equation for $\Phi$, viz., 
\begin{eqnarray}
\label{equation for Phi}
\frac{D^2 \Phi}{\Phi}  = - \frac{\left(D_{\alpha}\Phi\right)\left(D^{\alpha}\phi\right)}{\Phi \phi}.
\end{eqnarray}
Substituting this into $R^{(5)} = \gamma^{AB}R_{AB}$ we find 
\begin{eqnarray}
\label{R5 in terms of R4}
R^{(5)} =  R^{(4)}
   + \frac{2 \left(D_{\alpha}\Phi\right)\left(D^{\alpha}\phi\right)}{\Phi \phi},
\end{eqnarray}
where $R^{(4)} = g^{\alpha\beta}R_{\alpha\beta}^{(4)}$ is the scalar curvature of the spacetime hypersurfaces $\Sigma_{y}$.

We are now ready to obtain the effective equations for gravity in $4D$. With this aim we substitute the first equation in (\ref{dimensional reduction}) and (\ref{R5 in terms of R4}) into (\ref{spacetime components of G5}) and isolate $G_{\mu\nu}^{(4)} = R_{\mu\nu}^{(4)} - g_{\mu\nu} R^{(4)}/2$. The result can be written as
\begin{eqnarray}
\label{field equations for g}
G_{\mu\nu}^{(4)} =  \frac{8\pi}{\phi} T_{\mu\nu}^{(BD)} 
 + \frac{\omega}{\phi^2}\left[\left(D_{\mu}\phi\right) \left(D_{\nu}\phi\right) - \frac{1}{2}g_{\mu\nu} \left(D_{\alpha}\phi\right)\left(D^{\alpha}\phi\right)\right] + \frac{1}{\phi}\left(D_{\mu}D_{\nu}\phi - g_{\mu\nu} D^2\phi\right) - g_{\mu\nu}\frac{V(\phi)}{2 \phi}
\end{eqnarray}
where we have introduced the quantity  $V(\phi)$, which (as we will see bellow) plays the role of an effective or induced scalar potential;
and $T_{\mu\nu}^{(BD)}$ can be interpreted as an induced  EMT for an effective BD theory in $4D$. It is given by
\begin{eqnarray}
\label{T(BD)}
\frac{8\pi}{\phi}T_{\mu\nu}^{(BD)} &\equiv& \frac{D_{\mu}D_{\nu}\Phi}{\Phi} + \frac{g_{\mu\nu}V}{2 \phi}.
\end{eqnarray}
Taking the trace of  (\ref{field equations for g}) we   obtain a simple relation between  $R^{(4)}$ and  $T^{(BD)} = g^{\mu\nu} T_{\mu\nu}^{(BD)}$, namely  
  \begin{eqnarray}
\label{R4 in terms of TBD}
R^{(4)} = - \frac{8\pi}{\phi}T^{(BD)} +  \frac{\omega \left(D_{\alpha}\phi\right)\left(D^{\alpha}\phi\right)}{\phi^2} + \frac{3 D^2 \phi}{\phi} + \frac{2 V}{\phi}.
\end{eqnarray}

To obtain the equation of motion of $\phi$ in $4D$,  we substitute   (\ref{R5 in terms of R4}) and (\ref{R4 in terms of TBD}) into (\ref{dalambertian in 5D in terms of R5}). After some manipulations we get    

\begin{equation}
\label{D2 phi with residual term}
D^2\phi = \frac{8\pi}{3 + 2\omega}  T^{(BD)} + \frac{1}{3 + 2 \omega}\left[\phi \frac{d V(\phi)}{d \phi} - 2 V(\phi)\right],
\end{equation}
where the potential is derived from the equation
\begin{eqnarray}
\label{residual term}
\phi \frac{dV(\phi)}{d \phi} \equiv  &-& {2\left(1 + \omega\right)}\frac{\left(D_{\alpha}\Phi\right)\left(D^{\alpha}\phi\right)}{\Phi}.
\end{eqnarray}
The above equations constitute the basis for our  discussion. To an observer in $4D$, who is not aware of the existence of an extra dimension,  (\ref{field equations for g}) and (\ref{D2 phi with residual term}) are nothing but the   Brans-Dicke FE  in $4D$ with a self interacting potential $V$ and an effective EMT given by (\ref{T(BD)}). We note the importance of the potential: If $V = 0$, then the  effective FE in $4D$  
only yield the  BD theory with parameter\footnote{It is interesting to note that some string theories in the low energy limit also reduce to BD theory with $\omega = - 1$  \cite{Dabrowski}.} $\omega = - 1$.  Thus, although the introduction of $V(\phi)$ in (\ref{field equations for g}) might at first glance  look  artificial, $V \neq 0$ is necessary to obtain a general BD theory in $4D$ (a more detailed discussion is provided at the end of section $4$). In what follows, to avoid the so-called ghost fields (fields with the ``wrong" sign of the kinetic term),   we assume $\omega > - 3/2$.

\section{Brans-Dicke cosmology in $5D$}

In cosmological applications, under the  assumption of  spatial isotropy and homogeneity,  the line element in $5D$  is taken to be an extended version of the conventional Friedmann-Robertson-Walker metric in $4D$, namely
\begin{equation}
\label{cosmological metric in 5D, with y dependence}
dS^2 = dt^2 - a^2(t)\left[\frac{dr^2}{1 - k r^2} + r^2 \left(d\theta^2 + \sin^2 \theta d\varphi^2\right)\right] + \epsilon \Phi^2(t)dy^2,
\end{equation}
where $k = 0, + 1, - 1$ and $(t, r, \theta, \phi)$ are the usual coordinates for a spacetime with spherically symmetric spatial sections. 
For this line element the vacuum  $(T_{AB}^{(5)} = 0) $ Brans-Dicke field equations in $5D$ reduce as follows.  
From the $5D$ wave equation (\ref{equation for phi}) we obtain
\begin{equation}
\label{Dalambertian in 5D }
\ddot{\phi} + \dot{\phi}\left(\frac{3 \dot{a}}{a} + \frac{\dot{\Phi}}{\Phi}\right) = 0.
\end{equation}
Using this expression in (\ref{equation for the metric in 5D}), the temporal component $A = B = 0$ becomes
\begin{eqnarray}
\label{FE(00)}
3\frac{\dot{a}}{a}\left(\frac{{\dot{a}}}{a} + \frac{\dot{\Phi}}{ \Phi}\right) + \frac{3 k }{a^2}  = \frac{1}{\phi}\left(\ddot{\phi} + \frac{\omega {\dot{\phi}}^2}{2 \phi} \right). 
\end{eqnarray}
Again using (\ref{Dalambertian in 5D }), the spatial components $A = B = 1, 2, 3$  give
\begin{eqnarray}
\label{FE(11)}
\frac{2\ddot{a}}{a} + \frac{\dot{a}}{a}\left(\frac{\dot{a}}{a} + \frac{2\dot{\Phi}}{\Phi}\right) + \frac{\ddot{\Phi}}{\Phi}  + \frac{k }{a^2} 
= \frac{\dot{\phi}}{ \phi}\left(\frac{\dot{a}}{a} - \frac{\omega \dot{\phi}}{2 \phi}\right). 
\end{eqnarray}
Similarly, the $A = B = 4$ component yields
\begin{eqnarray}
\label{FE(44)}
3\left[\frac{\ddot{a}}{a} + \left(\frac{\dot{a}}{a} \right)^2\right] + \frac{3 k}{a^2}  = 
\frac{\dot{\phi}}{\phi}\left(\frac{\dot{\Phi}}{\Phi} - \frac{\omega \dot{\phi}}{2 \phi}\right). 
\end{eqnarray}

After a simple integration, from (\ref{Dalambertian in 5D }) we obtain
\begin{equation}
\label{integrating Dalambertian in 5D}
\dot{\phi}a^3\Phi = c_{1} = \mbox{constant} \neq 0.
\end{equation}

A similar expression, with $\phi \leftrightarrow \Phi$,  is obtained from (\ref{FE(00)})-(\ref{FE(44)}) if  we proceed as follows: (i) isolate $k$ from (\ref{FE(11)}); (ii) introduce  $k$ into (\ref{FE(00)}) and isolate for $\ddot{a}$; (iii) do the same for  (\ref{FE(44)}); (iv) equate the expressions for $\ddot{a}$ obtained in the previous two steps; (iv) use (\ref{Dalambertian in 5D }) to get rid of $\ddot{\phi}$. The result is a simple second-order differential equation for $\Phi$ whose first integral is given by
\begin{equation}
\label{Phi neq constant}
\dot{\Phi}a^3 \phi = c_{2} = \mbox{constant} \neq 0.
\end{equation}
(We assume  $\Phi \neq $ constant, otherwise the reduced $4D$ spacetime is empty, i.e., $T_{\mu\nu}^{(BD)} = 0$.)
Combining (\ref{integrating Dalambertian in 5D}) and (\ref{Phi neq constant}) we find 

\begin{equation}
\label{Phi in terms of phi}
\Phi = c_{3} \phi^{c_{2}/c_{1}},
\end{equation}
where $c_{3}$ is a constant of integration. 
Thus, if we know $a(t)$, we can construct the solution. Namely, we obtain $\phi$ by integrating $c_{3}\phi^{c_{2}/c_{1}}\dot{\phi} = c_{1}/a^3 $ and then $\Phi$ is given by (\ref{Phi in terms of phi}).

To find the  equation of motion for $a(t)$  we add (\ref{FE(00)}) and (\ref{FE(44)}). Then, we eliminate $\Phi$ and $\dot{\phi}$ by using    (\ref{integrating Dalambertian in 5D}) and (\ref{Phi neq constant}). We obtain 
\begin{equation}
\label{equation 1 for phi in terms of a two dots and adot}
\left(c_{1} + c_{2}\right)\dot{a} \phi^{- \left(c_{1} + c_{2}\right)/c_{1}} + c_{3} a^2  \left(2 {\dot{a}}^2 + 2 k + a \ddot{a}\right) = 0.
\end{equation}
The same process applied to (\ref{FE(44)}) yields

\begin{equation}
\label{equation 2 for phi in terms of a two dots and adot}
 c_{1}\left(\omega c_{1} - 2 c_{2}\right)\phi^{- 2 \left(c_{1} + c_{2}\right)/ c_{1}}  + {6  c_{3}^2 a^4 \left({\dot{a}}^2 +  k + a \ddot{a}\right)} = 0.
\end{equation}
The last two expressions generate the required equation for $a(t)$, namely\footnote{For a detailed study of the space of solutions  (which is not the object of this paper), it is convenient to write (\ref{equation for the scale factor}) in the form
$\alpha \left(\frac{dQ}{d z}\right)^2 + 4\left(\beta Q + k \alpha\right)\left(\frac{d Q}{dz} + Q + k\right) = 0$, where $Q \equiv \dot{a}^2$ 
and $z \equiv 2\ln{a}$.} 
\begin{equation}
\label{equation for the scale factor}
\alpha a^2 {\ddot{a}}^2 + 4\left(\beta {\dot{a}}^2 + k\alpha\right)\left(a \ddot{a} + {\dot{a}}^2 + k\right) = 0,
\end{equation}
where
\begin{equation}
\alpha = c_{1}\left(\omega c_{1} - 2 c_{2}\right), \;\;\;\beta = \left(\omega + \frac{3}{2}\right)c_{1}^2 + c_{1}c_{2} + \frac{3 c_{2}^2}{2}.
\end{equation}
It should be noted that the field equation (\ref{FE(11)}) is identically satisfied by (\ref{equation 1 for phi in terms of a two dots and adot})-(\ref{equation 2 for phi in terms of a two dots and adot}). Therefore, {\it any} solution to (\ref{equation for the scale factor}) gives rise to  an exact solution to the FE (\ref{Dalambertian in 5D })-(\ref{FE(44)}).

\medskip
For $k \neq 0$ the solutions to (\ref{equation for the scale factor}) can be expressed in terms of elementary functions only in some particular cases, which  correspond to specific choices of the constants, e.g., $c_{1} = - c_{2}$, $\alpha = 0$, $\beta = 0$, etc.  However, for $k = 0$ we find that (\ref{equation for the scale factor}) admits a unique solution, which is
\begin{equation}
\label{general solution with k = 0}
a(t) = \left(C_{1} t + C_{2}\right)^{l},
\end{equation}
where $C_{1}$ and $C_{2}$ are constants of integration and $l$ is a parameter that depends (in a very complicated way) on $c_{1}$, $c_{2}$ and $\omega$. We note that astrophysical data from WMAP \cite{Spergel}
and BOOMERANG \cite{BOOMERANG}, analyzed in the context of models based on GR, indicate that $k = 0$ around the present epoch. However, any other theory could give in principle  very  different answer.

\medskip

In this work we restrict our attention to the BD cosmological models generated by the exact solution (\ref{general solution with k = 0}).  From a practical viewpoint,  the solution looks much simpler in terms of the new parameters $l$ and $m$ (instead of $c_{1}$ and $c_{2}$), namely

\begin{eqnarray}
\label{Case 3,  solution 2}
dS^2 &=& dt^2 - B^2 t^{2 l}\left[d r^2 + r^2 \left(d\theta^2 + \sin^2\theta d\varphi^2\right)\right] + \epsilon C^2 t^{2 m}d y^2,\nonumber \\
\phi &=& D t^{\left(1 - m - 3 l\right)}, \;\;\; \omega = - \frac{12 l^2 + 2\left(m - 1\right)\left(3 l + m\right)}{\left(1 - m - 3 l\right)^2}. 
\end{eqnarray}
 
We note that $\phi =$ constant and $\omega \rightarrow \infty$ in the limit $m \rightarrow \left(1 - 3 l\right)$. In this limit, the field equations can only\footnote{We disregard the ``static" case where $l = 0$.} be satisfied if $l = 1/2$, i.e., $m = - 1/2$. Therefore, we recover the $5D$ general-relativistic solution
\begin{eqnarray}
\label{general-relativistic solution}
dS^2 &=& dt^2 - B^2 t \left[d r^2 + r^2 \left(d\theta^2 + \sin^2\theta d\varphi^2\right)\right] + \frac{\epsilon C^2}{t} d y^2, 
\end{eqnarray}
which is the unique solution to the Einstein field equations $G_{AB} = 0$ in $5D$ for the metric (\ref{cosmological metric in 5D, with y dependence}) with flat spatial sections $(k = 0)$.

\section{Effective Brans-Dicke cosmology in $4D$}

We now proceed to study the effective $4D$ picture generated by the spatially flat $5D$ solutions discussed in the preceding section. 
For the line element (\ref{cosmological metric in 5D, with y dependence}),  the non-vanishing components of the induced Brans-Dicke energy-momentum tensor (\ref{T(BD)}) are
\begin{eqnarray}
\label{components of the Brans-Dicke theory}
\frac{8\pi}{\phi} {T_{0}^{0}}^{(BD)} &=& \frac{\ddot{\Phi}}{\Phi} + \frac{V}{2 \phi},\nonumber \\
\frac{8 \pi}{\phi}{T_{1}^{1}}^{(BD)} &=& \frac{\dot{a}\dot {\Phi}}{a \Phi}  + \frac{V}{2 \phi},
\end{eqnarray}
where $V = V(\phi)$ should be determined from (\ref{residual term}). We note that ${T_{2}^{2}}^{(BD)} = {T_{3}^{3}}^{(BD)} = {T_{1}^{1}}^{(BD)}$, which means that the induced EMT looks like a perfect fluid with energy density $\rho = {T_{0}^{0}}^{(BD)}$ and isotropic pressure $p = - {T_{1}^{1}}^{(BD)}$.

We emphasize  that as in the conventional $4D$ Brans-Dicke theory,  in our models the (effective) EMT obeys the ordinary  conservation law $D_{\mu} {T^{\mu}_{\nu}}^{(BD)} = 0$ (the same as in Einstein's theory), which in the cosmological realm yields  the usual equation of motion 
\begin{equation}
\label{conservation equation}
\dot{\rho} + 3\frac{\dot{a}}{a}\left(\rho + p\right) = 0.
\end{equation}

Substituting (\ref{Case 3,  solution 2}) into (\ref{residual term}) we obtain
\begin{equation}
\phi \left(\frac{d V}{d\phi}\right) = - \frac{2 m \left(m^2 + 3 l^2 - 1\right)}{\left(m + 3 l - 1\right) D^{2/\left(m + 3 l - 1\right)}}\phi^{\left(m + 3 l + 1\right)/\left(m + 3 l - 1\right)}.
\end{equation}
Integrating this equation, and setting the integration constant equal to zero, we find 
\begin{equation}
\label{V}
V(\phi) = - \frac{2 m \left(m^2 + 3 l^2 - 1\right)}{\left(m +  3 l +1 \right) D^{2/\left(m + 3 l - 1\right)}}\phi^{\left(m + 3 l + 1\right)/\left(m + 3 l - 1\right)}.
\end{equation}
The effective EMT  (\ref{components of the Brans-Dicke theory}) is given by  
\begin{eqnarray}
\label{TBD for cosmologies from  solutions independent of y and V non-zero}
\frac{8 \pi}{\phi}{T_{0}^{0}}^{(BD)} &=&  \frac{3 m l \left(m - l - 1\right)}{t^2 \left(m + 3 l + 1\right)},\nonumber \\
\frac{8 \pi}{\phi}{T_{1}^{1}}^{(BD)} &=& - \frac{m \left(m + 1\right)\left(m - l - 1\right)}{t^2\left(m + 3 l + 1\right)}. 
\end{eqnarray}
Consequently, the equation of state is 
\begin{equation}
\label{introduction of n}
p = n \rho, \;\;\; n = \frac{m + 1}{3 l}.
\end{equation}
The parameters $m$ and $l$ can be expressed in terms of  $n$ and the acceleration parameter $q = - a \ddot{a}/{\dot{a}}^2$ as 
\begin{equation}
\label{m and l in terms of n and q}
m  = \frac{3 n}{q + 1} - 1, \;\;\;l = \frac{1}{q + 1}.  
\end{equation}
Since the present epoch of the universe is matter dominated we set $m = - 1$. Thus
\begin{eqnarray}
\frac{8 \pi}{\phi}{T_{0}^{0}}^{(BD)} = \frac{l + 2}{t^2}, \;\;\; \omega = - \frac{4\left[ 3l \left(l - 1\right) + 1\right]}{\left(2 - 3 l\right)^2}, \;\;\;V = \frac{2 l \phi^{3 l/\left(3 l - 2\right)}}{D^{2/(3 l - 2)}}.
\end{eqnarray}
For $n = 0$, the condition $\omega > - 3/2$ restricts the range of $l$ to be  either $l < 2\left(1 - \sqrt{2/3}\right) \approx 0.37$ 
or $l > 2\left(1 + \sqrt{2/3}\right) \approx 3.633$. The former range requires $q > \left(1 + \sqrt{6}\right)/2 \approx 1.72$, which is inapplicable to the present epoch. However, the latter range leads to accelerated cosmic expansion with 
 $q <  \left(1 - \sqrt{6}\right)/2 \approx - 0.72$. Thus, our reduced BD cosmological model expands in a way consistent with  current observed measurements $q = - 0.67 \pm 0.25$ \cite{Wendy} in the context of models based on GR. Besides, for dust $(m = -1)$ the extra dimension contracts while the spatial dimensions expand. 
A similar analysis can be easily done for any value of $n$. 

In summary,  the reduced Brans-Dicke  cosmological solution in $4D$, which is obtained from (\ref{Case 3,  solution 2}), can be written  as

\begin{eqnarray}
\label{Eq. for the summary 1}
ds^2 &\equiv& dS^2_{|\Sigma_{y}} = dt^2 - B^2 t^{2 l}\left[dr^2+ r^2 \left(d\theta^2 + \sin^2 \theta d\varphi^2\right)\right],\nonumber \\
\rho &=&  \rho_{0} \left(\frac{t_{0}}{t}\right)^{3 l (n + 1)}, \nonumber \\
\phi &=& \phi_{0} \left(\frac{t}{t_{0}}\right)^{[2 - 3 l\left(n + 1\right)]},\nonumber \\
V &=& V_{0}\left(\frac{\phi}{\phi_{0}}\right)^{3 l\left(n + 1\right)/\left[3 l\left(n + 1\right) - 2\right]},
 \end{eqnarray}
with 
\begin{eqnarray}
\label{Eq. for the summary 1. The constants}
\phi_{0} &=& \frac{8 \pi \left(n + 1\right) \rho_{0} t_{0}^2}{\left(l + 2 - 3 l n\right)\left(1 - 3 l n\right)}, \nonumber \\
V_{0} &=& \frac{16 \pi \rho_{0}\left[l \left(1 + 3 n^2\right) - 2 n\right]}{ 2 + l\left(1 - 3 n\right) },   
\end{eqnarray}
where $\rho_{0}$ refers to the value of the energy density at some arbitrary fixed time $t_{0}$,  
and 
\begin{equation}
\label{l for the summary}
l = l_{(\pm)} = \frac{6\left(\omega + 1\right) + 3 n\left(2\omega + 3\right) \pm \sqrt{9 n^2 + 12 \left(1 + \omega\right)\left(3 n - 1\right)}}{9\left(\omega + 2\right)n^2 + 18\left(\omega + 1\right) n + 3\left(3 \omega + 4\right)}.
\end{equation}
It is worth noticing that regardless of $n$ for large values of $|\omega|$  we get $l = 2/3\left(n + 1\right)$, which implies $V = 0$ and $\phi = $ constant. Also, in this limit from (\ref{introduction of n}) it follows that $m = \left(n - 1\right)/\left(n + 1\right)$, which means that the extra dimension contracts, or compactifies, for any $- 1 < n < 1$. Consequently,  formally for $|\omega| \rightarrow \infty$ we recover the  usual spatially flat FRW cosmology of ordinary general relativity.

Since the term inside of the root must be non-negative, we find $\omega < - 1$ for dust. However, for radiation  there are no restrictions on $\omega$. For $n = 1/3$  we find $l = l_{(+)} = 1/2$, which is identical to the radiation-dominated epoch of general relativity, 
regardless of the specific value of $\omega$. Therefore, the effective BD cosmology in $4D$  can give decelerated radiation era as well as accelerating matter dominated era.

\medskip

 For the sake of comparison, we note that the well-known spatially flat BD dust solutions \cite{Dicke} exist for any value of $\omega$ and  show accelerated expansion only 
for $- 2 \leq \omega \leq - 3/2$ (see \cite{Pavon} and references therein). However, our dust model requires $\omega < -1$ and the accelerated expansion  occurs  in the range $-3/2 < \omega < - 1$, where the lower limit comes from the positive energy condition on the scalar field. The simultaneous occurrence of $\omega > - 3/2$ and accelerated expansion is a consequence  of the non-vanishing scalar potential.

\medskip 

If we assume $V = 0$ at the beginning, then we obtain a very limited class of BD cosmologies in $4D$. In fact, in  the case where $V = 0$ 
 instead of (\ref{TBD for cosmologies from  solutions independent of y and V non-zero}) we have 
\begin{eqnarray}
\label{TBD for cosmologies from  solutions independent of y and V = 0}
\frac{8 \pi}{\phi}{T_{0}^{0}}^{(BD)} &=&  \frac{m\left(m - 1\right)}{t^2},\nonumber \\
\frac{8 \pi}{\phi}{T_{1}^{1}}^{(BD)} &=& \frac{l m}{t^2}, 
\end{eqnarray}
Besides, from (\ref{residual term}) it follows that $\omega = -1$. Then, from (\ref{Case 3,  solution 2}) we obtain $m = \pm \sqrt{1 - 3 l^2}$, which requires $|l| \leq 1/\sqrt{3} \approx 0.577$. The choice  $m = - \sqrt{1 - 3 l^2}$ assures that $\rho > 0 $ and $p > 0$. When $l = 1/2$ $(m = - 1/2)$  we have $\phi =$ constant and recover the spatially flat FRW solution of 
general relativity with $p = \rho/3$, which is the spacetime section of (\ref{general-relativistic solution}).

\section{$5D$ scalar-vacuum Brans-Dicke theory with variable $\omega$}

We have seen that the scalar-vacuum Brans-Dicke cosmology in $5D$ yields accelerated expansion of an effective matter-dominated $4D$ universe if $- 3/2 < \omega < - 1$. This range of $\omega$ is consistent with the one obtained from the reduced BD cosmologies with higher-dimensional matter discussed in \cite{Li}. However, these  small values of $|\omega|$  are in sharp contradiction with  the solar system bound $\omega > 600$. 
This contradiction is consistently found when the ordinary BD theory  in $4D$ is applied to  cosmological problems like inflation and structure formation, not only to the late time cosmic acceleration (see e.g. \cite{Sen} and references therein). 
As a possible  way out of this problem, it  has been suggested to consider a modified version of BD theory where the parameter $\omega$ is a function of the scalar field rather than a constant \cite{Pavon}.  

In this section we study the question of whether accelerated expansion is compatible with large values of $\omega$, within the context of a   $5D$ scalar-vacuum BD theory with variable $\omega$. With this aim we integrate the FE and obtain an explicit equation relating $q$ and $\omega$. Our analysis shows that the answer to our question is negative, namely, late time accelerated expansion is incompatible with large values of $\omega$.

For a  variable $\omega$  the wave  equation (\ref{equation for phi}) becomes

\begin{equation}
\label{wave equation with varying omega}
\nabla^2 \phi = - \frac{3}{2\left(3\omega + 4\right)}\frac{d \omega}{d \phi} \left(\nabla_{A}\phi\right)\left(\nabla^A \phi\right).
\end{equation}
Therefore, instead of (\ref{Dalambertian in 5D }) we now have
\begin{equation}
\label{Dalambertian in 5D with varying omega}
\ddot{\phi} + \dot{\phi}\left(\frac{3 \dot{a}}{a} + \frac{\dot{\Phi}}{\Phi}\right) = - \frac{3 \dot{\omega} \dot{\phi}}{2\left(3\omega + 4\right)}.
\end{equation}
This equation gives the first integral 
\begin{equation}
\label{integrating Dalambertian in 5D with varying omega}
\dot{\phi}a^3\Phi = {\bar{c}}_{1}/ \sqrt{3\omega + 4},
\end{equation}
where ${\bar{c}}_{1}$ is a constant of integration. In practice this is a definition of $\omega$ in terms of the metric functions and the scalar field. It implies  that $\omega$ should obey the condition $\omega \geq - 4/3$. In this regard we note that no such condition follows from the FE when $\omega$ is constant.

Instead of the simplified FE (\ref{FE(00)})-(\ref{FE(44)}) we now have

\begin{eqnarray}
\label{general FE(00)}
3\frac{\dot{a}}{a}\left(\frac{{\dot{a}}}{a} + \frac{\dot{\Phi}}{ \Phi}\right) + \frac{3 k }{a^2}  = - \frac{\dot{\phi}}{\phi}\left( \frac{3 \dot{a}}{a} + \frac{\dot{\Phi}}{\Phi} -  \frac{\omega {\dot{\phi}}}{2 \phi} \right). 
\end{eqnarray}

\begin{eqnarray}
\label{general FE(11)}
\frac{2\ddot{a}}{a} + \frac{\dot{a}}{a}\left(\frac{\dot{a}}{a} + \frac{2\dot{\Phi}}{\Phi}\right) + \frac{\ddot{\Phi}}{\Phi}  + \frac{k }{a^2} 
=  - \frac{\ddot{\phi}}{\phi} -  \frac{\dot{\phi}}{ \phi}\left( \frac{2 \dot{a}}{a} + \frac{\dot{\Phi}}{\Phi} + \frac{\omega \dot{\phi}}{2 \phi}\right). 
\end{eqnarray}

\begin{eqnarray}
\label{general FE(44)}
3\left[\frac{\ddot{a}}{a} + \left(\frac{\dot{a}}{a} \right)^2\right] + \frac{3 k}{a^2}  = 
- \frac{\ddot{\phi}}{\phi}  - \frac{\dot{\phi}}{\phi}\left(\frac{3 \dot{a}}{a} + \frac{\omega \dot{\phi}}{2 \phi}\right). 
\end{eqnarray}

From (\ref{general FE(00)}) and (\ref{general FE(44)}) we eliminate  $\omega$ to obtain a first-order differential equation for $\Phi$.  This equation admits exact integration for  $k = 0$, namely 
\begin{equation}
\label{nice first integral 1}
\Phi a^2\left(3\dot{a}\phi + \dot{\phi} a\right) = {\bar{c}}_{2}.
\end{equation}
Similarly, from (\ref{general FE(11)}) and (\ref{general FE(44)}) we get a first-order differential equation for $\phi$,  which for $k = 0$ yields 
\begin{equation}
\label{nice first integral 2}
\phi a^2\left(\dot{a}\Phi  - \dot{\Phi} a \right) = {\bar{c}}_{3}.
\end{equation}
In the above expressions ${\bar{c}}_{2}$ and ${\bar{c}}_{3}$ are constants of integration. 

An important result follows from  $a^3\left(3\phi \dot{\Phi} + \Phi \dot{\phi}\right) = {\bar{c}}_{2} - 3 {\bar{c}}_{3}$ after substituting (\ref{integrating Dalambertian in 5D with varying omega}) into it. Specifically,  

\begin{equation}
\label{Phi dot for varying omega}
\dot{\Phi} a^3 \phi = - \frac{{\bar{c}}_{1}}{3 \sqrt{3 \omega + 4}} + \frac{{\bar{c}}_{2}}{3} - {\bar{c}}_{3}.
\end{equation}
This equation says that a varying $\omega$ demands $\dot{\Phi} \neq 0$, otherwise $\omega = $ constant.

Equation (\ref{nice first integral 1})  divided by (\ref{nice first integral 2}) yields a first-order differential equation which can be easily integrated to  find
\begin{eqnarray}
\label{general case}
\phi  &\propto& \Phi^{- {\bar{c}}_{2}/{\bar{c}}_{3}} a^{\left({\bar{c}}_{2} - 3 {\bar{c}}_{3}\right)/{\bar{c}}_{3}}, \;\;\;{\bar{c}}_{3} \neq 0,\nonumber \\
\Phi &\propto&  a, \;\;\;{\bar{c}}_{3} = 0. 
\end{eqnarray}
Substituting these expressions back into either (\ref{nice first integral 1}) or (\ref{nice first integral 2}), and using  (\ref{general FE(00)})-(\ref{general FE(44)}),  we obtain the general expressions for   $\Phi$, $\phi$ and $\omega$ in terms $a(t)$. There are three families of solutions listed bellow as I-III. 
\paragraph{I:} 
\begin{eqnarray}
\label{I}
\Phi &=& A a F^{{\bar{c}}_{3}/\left({\bar{c}}_{3} -  {\bar{c}}_{2}\right)}, \;\;\; {\bar{c}}_{2} - {\bar{c}}_{3} \neq 0\nonumber\\
\phi &=& B  a^{- 3} F^{{\bar{c}}_{2}/\left({\bar{c}}_{2} - {\bar{c}}_{3}\right)},\nonumber \\
\omega &=& - \frac{2\left(6 F^2 \dot{a}^2 - 4 {\bar{c}}_{2} F \dot{a} + {\bar{c}}_{2} {\bar{c}}_{3}\right)}{\left(3 F \dot{a} - {\bar{c}}_{2}\right)^2}, 
\end{eqnarray}
where $F$ is given by 
\begin{equation}
\label{definition of F}
F \equiv C  + \left({\bar{c}}_{2} - {\bar{c}}_{3}\right)f(t), \;\;\;\; f(t) \equiv \int{\frac{d t}{a(t)}}, 
\end{equation}
and $A$, $B$, $C$ are arbitrary constants. 
\paragraph{II:}
\begin{eqnarray}
\label{II}
\Phi &=& A a e^{- {\bar{c}}_{2}  f/C}, \;\;\; {\bar{c}}_{2} - {\bar{c}}_{3} = 0,\;\;\; {\bar{c}}_{2} = {\bar{c}}_{3} \neq 0,\nonumber \\
\phi &=& (C/A)  a^{- 3} e^{{\bar{c}}_{2}  f/C},  \nonumber \\
\omega &=& - \frac{2\left(6 C^2 \dot{a}^2 - 4 {\bar{c}}_{2} C \dot{a} + {\bar{c}}_{2}^2\right)}{\left(3 C \dot{a} - {\bar{c}}_{2}\right)^2}.
\end{eqnarray}

\paragraph{III:}
\begin{eqnarray}
\label{III}
\Phi &=&  A a,  \;\;\;  {\bar{c}}_{2} =  {\bar{c}}_{3} = 0,\nonumber \\
\phi &=&  B a^{- 3},  \nonumber \\
\omega &=& - 4/3.
\end{eqnarray}
One can verify that (\ref{I})-(\ref{III}) satisfy the field equations (\ref{general FE(00)})-(\ref{general FE(44)}), as expected.

$\bullet$ We now substitute the general solution (\ref{I}) into the ``new" wave equation (\ref{Dalambertian in 5D with varying omega}). The resulting  equation allows us to express  the deceleration parameter  $q = - a\ddot{a}/\dot{a}^2$ as a function of $\omega$. For this we use

\begin{eqnarray}
\label{a two dots and a dot}
\ddot{a} &=& - q \frac{\dot{a}^2}{a}, \nonumber \\
\dot{a} &=& {\dot{a}}_{(\pm)} = \frac{{\bar{c}}_{2}\left[\sqrt{3\omega + 4} \pm \sqrt{2\left(2 - 3\alpha\right)}\right]}{3 F \sqrt{3\omega + 4} }, \;\;\;\alpha \equiv \frac{{\bar{c}}_{3}}{{\bar{c}}_{2}} \leq \frac{2}{3},
\end{eqnarray}
where the second equation comes from the expression for $\omega$ in (\ref{I}),  and $\alpha$ is a dimensionless parameter. It should be noted that ${\dot{a}}_{(+)}$ represents an ever expanding (contracting) universe, while ${\dot{a}}_{(-)}$ changes its motion/sign for $\omega = - 2 \alpha$.

After a long but straightforward calculation we find\footnote{We observe that these expressions are obtained without the explicit knowledge of the function $F$ (or $f$) introduced in (\ref{definition of F}). }
\begin{eqnarray}
\label{expression for q}
q_{(+)}(\omega, \alpha) &=& \frac{3\sqrt{3\omega + 4}\left(1 - \alpha\right)\left[3\omega + 4  + \sqrt{3\omega + 4}\left(\sqrt{4 - 6\alpha} - 1\right) - \sqrt{4 - 6\alpha}\right]}{\left(3\omega + 4\right)\left(2\sqrt{4 - 6\alpha} - 1\right) - \left(4 - 6\alpha\right) + \sqrt{3\omega + 4}\left[3\omega + 4 - 2\sqrt{4 - 6\alpha} + \left(4 - 6\alpha\right)\right]}, \nonumber \\
\\
q_{(-)}(\omega, \alpha) &=& \frac{3\sqrt{3\omega + 4}\left(1 - \alpha\right)\left[- \left(3\omega + 4\right)  + \sqrt{3\omega + 4}\left(\sqrt{4 - 6\alpha} + 1\right) - \sqrt{4 - 6\alpha}\right]}{\left(3\omega + 4\right)\left(2\sqrt{4 - 6\alpha} + 1\right) + \left(4 - 6\alpha\right) -  \sqrt{3\omega + 4}\left[3\omega + 4 + 2\sqrt{4 - 6\alpha} + \left(4 - 6\alpha\right)\right]},
\end{eqnarray}
where $q_{(+)}$ and $q_{(-)}$ correspond to the choice of positive or negative sign in (\ref{a two dots and a dot}), respectively. They are equal to each other only for $\omega = - 4/3$ $(q_{(\pm)} = 0)$; $\alpha = 2/3$ ($q_{(\pm)} = 1$),   and in the limit $\omega \to \infty$, namely
\begin{equation}
\label{limitting value of q}
\lim_{\omega \to \infty}{q_{(\pm)}} = 3\left(1 - \alpha\right) \geq 1,
\end{equation}
which follows from the requirement $\alpha \leq 2/3$. The denominator of $q_{(\pm)}$ vanishes at $\omega = -1$ and $\omega = - 2\alpha$. However, it is not difficult to verify that $q_{(+)}$ remains finite and positive for all values of $\omega$ and $\alpha$. In addition,  for every $\alpha \leq 2/3$ we find that $q_{(+)}$ increases monotonically with $\omega$ from zero at $\omega = - 4/3$ to $\left[3\left(1 - \alpha\right)\right]$ as $\omega \to \infty$.
  
The model generated by  ${\dot{a}}_{(-)}$ is regular at $\omega = - 1$. It represents a universe that  expands (contracts) for $\omega < - 2\alpha$ and reverses its cycle for $\omega > - 2\alpha$ in such a way that $q_{(-)}$ diverges at $\omega = - 2\alpha$. Indeed,  $q_{(-)}$ goes monotonically from $0$ to $- \infty$ as $\omega$ increases from $\omega = - 4/3$ to $\omega = - 2\alpha$. Conversely,   $q_{(-)}$ goes monotonically from $+ \infty$ to $\left[3\left(1 - \alpha\right)\right]$ as $\omega$ increases from $\omega = - 2 \alpha$ to $\omega \to \infty$.

Thus, in the framework under consideration, in this section we have used the general spatially-flat solution (\ref{I}) to demonstrate  that accelerated expansion $(q < 0)$ in an (ever)expanding universe is  incompatible with large positive values of $\omega$. The question of whether this result still holds for $k \neq 0$ remains open.

\newpage

\section{Summary} 

In this work we have investigated the question of whether,  we can get the observed  late time  accelerated universe from a scalar-vacuum Brans-Dicke theory in $5D$. Thus, we have not introduced matter fields in $5D$. Rather, after dimensional reduction we have seen that   the effective equations in $4D$ can be regarded as those for  the standard BD in $4D$ with a self interacting potential $V = V(\phi)$ and a matter field. These  appear  in $4D$ as a consequence of the  variation of $\Phi$ with time, i.e., $V = 0$ and $T_{\mu\nu}^{(BD)} = 0$ when $\Phi =$ constant.
In the case that  $\omega = - 1$, from (\ref{residual term}) it follows   that we can set $V= 0$ without loss of generality. However, if $\omega \neq - 1$ (and $\Phi \neq $ constant) we cannot  assume $V = 0$. Thus, in our work the induced potential is not picked by hand. Instead, it is dictated by the geometry in $5D$. 

As a consequence, we have shown that the $5D$ scalar-vacuum BD  solutions (\ref{Case 3,  solution 2}) can be interpreted by an observer in $4D$ as a  family of BD cosmological models (\ref{Eq. for the summary 1})-(\ref{l for the summary}) with matter and an effective potential. These models   can give decelerated radiation era as well as accelerating matter dominated era. Formally, in the limit $|\omega| \rightarrow \infty$ they reduce to  the spatially flat FRW cosmologies of general relativity.

Since small values of $|\omega|$  are in  contradiction with  the solar system bound $\omega > 600$, here
we have studied the question of whether accelerated expansion is compatible with large values of $\omega$, within the framework of a  $5D$ scalar-vacuum BD theory with {\it variable} $\omega$.  In this  framework, and using the  general class of  spatially flat solutions (\ref{I}), we have shown that the answer to this question is negative. Namely, accelerated expansion is  incompatible with large positive values of $\omega$. The origin of this incompatibility  may be the straightforward 
transportation of $\omega$ from large cosmological scales to the small scales of the solar system, without taking into consideration local inhomogeneities in astronomical scale. It is possible that such inhomogeneities can produce the large values of $\omega$ observed  in the solar system.

To finish the discussion, we should mention that, besides the obvious extension to cosmologies with $k \neq 0$,  our work leaves a number of important questions open for future research. 
For example,  we do not provide  a description of how and when the BD universe goes from decelerated to accelerated expansion. This is because the assumption that our $4D$ spacetime can be recovered by going onto a hypersurface $y = y_{0} = $ constant only leads to effective matter  that satisfies the barotropic equation of state (\ref{introduction of n}). To obtain more general forms of effective matter, which can allow  the study of the  evolution of $q$, we need a more ``flexible"  embedding approach\footnote{The effective spacetime measured by an observer depends on her/his state of motion. The simplest physical scenario emerges in the rest  (also called {\it comoving}) frame, which in the present case means $(d x^i = dy = 0)$. 
In such a frame, the spacetime is recovered by going onto some hypersurface $\Sigma_{y}:y = y_{0} =$ constant, which is  orthogonal to the unit 5D vector (\ref{unit vector n}).
A more ``flexible" approach, that respects the spatial homogeneity and isotropy of FRW models, is to consider that 4D observers are at rest only  in 3D $(dx^i = 0)$, but moving in 5D, i.e., that our spacetime  is recovered on a dynamical 4D hypersurface $y = y(t)$ or $t = t(\tau)$, $y = y(\tau)$ in parametric form \cite{JPdeL1}.}, e.g.,  the one considered in \cite{JPdeL1}. We have not  discussed how matter fields in $5D$ can affect of evolution of $\omega$, nor have we considered whether or not matter fields in $4D$ can be incorporated in the discussion. However, in the context of the present work, the matter content of spacetime is an effective result of the reduction from $5D$ to $4D$.
To add matter in $4D$ we should follow an approach similar to the one used in braneworld models, i.e., identify our spacetime with some singular hypersurface (the brane where ordinary $4D$ matter fields are located)  embedded in an empty $5D$ BD. This requires the consideration of more general solutions  where the $5D$ metric and the scalar field are functions of the extra coordinate,  
similar to the ones recently considered in \cite{Preprint}. Our simplified model shows that the scalar field in higher dimensional BD theory alone can be responsible for the present cosmic accelerated expansion, which rules out dark energy dynamics. We are currently working on constructing more realistic models.

\end{document}